\documentclass[11pt, a4paper, twocolumn]{article}

\usepackage[margin=0.75in, top=1in, bottom=0.9in, columnsep=0.3in]{geometry}
\usepackage{amsmath, amssymb}
\usepackage{graphicx}
\usepackage[numbers,sort&compress]{natbib}
\usepackage{booktabs}
\usepackage{float}
\usepackage{hyperref}
\usepackage{xcolor}
\usepackage{caption}
\usepackage{authblk}
\usepackage{enumitem}
\usepackage{microtype}
\usepackage{tabularx}
\usepackage{titlesec}
\usepackage{fancyhdr}

\definecolor{arxivblue}{RGB}{0, 51, 102}
\definecolor{linkblue}{RGB}{10, 82, 145}
\definecolor{citegreen}{RGB}{0, 100, 60}

\hypersetup{
    colorlinks=true,
    linkcolor=linkblue,
    citecolor=citegreen,
    urlcolor=linkblue
}

\titleformat{\section}{\large\bfseries}{\thesection.}{0.5em}{}
\titleformat{\subsection}{\normalsize\bfseries}{\thesubsection}{0.5em}{}
\titleformat{\subsubsection}{\normalsize\itshape}{\thesubsubsection}{0.5em}{}
\titlespacing*{\section}{0pt}{12pt plus 2pt minus 2pt}{6pt plus 1pt}
\titlespacing*{\subsection}{0pt}{8pt plus 2pt minus 1pt}{4pt plus 1pt}
\titlespacing*{\subsubsection}{0pt}{6pt plus 2pt minus 1pt}{3pt plus 1pt}

\fancypagestyle{plain}{
    \fancyhf{}
    \fancyhead[L]{\small\textit{Preprint}}
    \fancyhead[R]{\small\thepage}
    
}

\setlist{nosep, leftmargin=1.5em}

\title{\vspace{-1.5em}\textbf{\large Aircast-Mars: A Mars Foundation Model for Global Weather Forecasting with HEALPix-Aware Convolutions}\vspace{-0.5em}}

\author[1,2]{Manmeet Singh}
\author[3]{Saptarishi Dhanuka}
\author[1]{Naveen Sudharsan}
\author[4]{Houman Owhadi}
\author[1]{Krista M. Soderlund}
\author[5]{Alphan Altinok}

\affil[1]{\small The University of Texas at Austin, Austin, Texas, USA}
\affil[2]{\small Western Kentucky University, Bowling Green, Kentucky, USA}
\affil[3]{\small Ashoka University, Delhi-NCR, India}
\affil[4]{\small California Institute of Technology, Pasadena, California, USA}
\affil[5]{\small NASA Jet Propulsion Laboratory, Pasadena, California, USA}

\date{\vspace{-1em}\small May 2026}

\begin{document}

\twocolumn[
\maketitle
\vspace{-0.5em}
\begin{center}\textbf{Abstract}\end{center}
\vspace{-0.5em}
{\small\noindent
Foundation models for planetary atmospheres promise fast, lightweight surrogates of expensive general circulation models (GCMs) for mission planning and scientific inquiry. Here we present \textbf{Aircast-Mars}, a deep-learning weather prediction system for Mars trained on the Ensemble Mars Atmosphere Reanalysis System (EMARS) v1.0. We regrid temperature, zonal wind, and meridional wind fields across 28 vertical levels onto a hierarchical equal-area isolatitude pixelization (HEALPix) mesh at $N_\text{side} = 64$ (${\sim}110$~km resolution) and train a HEALPix-aware 2D U-Net inspired by the DLESyM architecture to predict the next hourly atmospheric state. The model employs custom inter-face padding that respects the topology of the 12-face HEALPix sphere and modern ConvNeXt residual blocks with capped Gaussian Error Linear Unit (GELU) activations. While containing 4.3 million trainable parameters—a compact size compared to terrestrial weather foundation models—the network achieves a best validation Mean Squared Error (MSE) of $1.58 \times 10^{-5}$ in normalized units. Recursive autoregressive rollouts remain stable and physically coherent for 25 hours (one Martian sol), with Root Mean Square Error (RMSE) growing monotonically from ${\sim}0.004$ at $t+1$\,h to ${\sim}0.031$ at $t+25$\,h without divergence. Compared to a baseline 3D U-Net, the HEALPix-aware architecture reduces validation loss by more than an order of magnitude while using fewer parameters. The model generates a one-hour forecast in approximately 0.5~seconds on a single GPU, offering several orders-of-magnitude speedup over traditional numerical GCMs. These results demonstrate that parsimonious, geometry-respecting neural architectures can capture synoptic-scale Martian atmospheric dynamics and provide a foundation for planetary-scale weather forecasting.\par}
\vspace{1em}
\vspace{1.5em}
]

\section{Introduction}

Mars' thin CO$_2$-dominated atmosphere supports a rich variety of dynamical phenomena---baroclinic waves, thermal tides, regional and global dust storms---that directly influence spacecraft operations~\cite{read2015physics}. Thermal tides, in particular, are strongly coupled to the dust cycle and can dominate the diurnal variability of the Martian atmosphere~\cite{leovy1979thermal}. Reliable medium-range forecasts could improve landing safety, rover route planning, and the scheduling of scientific observations. Yet traditional forecasting requires general circulation models (GCMs) and data assimilation schemes purpose-built for Mars, demanding substantial high-performance computing resources and significant wall-clock time~\cite{haberle2019}.

The Ensemble Mars Atmosphere Reanalysis System (EMARS) provides the best available observationally constrained reference dataset for the Martian atmosphere. EMARS v1.0 spans Mars years 24--33 (approximately 1999--2017) and assimilates Thermal Emission Spectrometer (TES) and Mars Climate Sounder (MCS) temperature retrievals into the Geophysical Fluid Dynamics Laboratory (GFDL) Mars GCM using the Local Ensemble Transform Kalman Filter (LETKF), producing hourly analyses on a $6^\circ$ longitude by $5^\circ$ latitude grid with 28 vertical levels~\cite{greybush2019}. The dataset provides temperature, zonal and meridional winds, surface pressure, and several aerosol tracers. These data have been used to study transient eddies, polar vortices, and dust storms~\cite{greybush2012ensemble}.

Parallel developments on Earth demonstrate that machine learning can emulate numerical weather models at a fraction of their cost. Karlbauer et al.~\cite{karlbauer2024} showed that a parsimonious deep-learning model forecasting only seven atmospheric variables on a ${\sim}110$~km HEALPix mesh~\cite{gorski2005healpix} with a 3-hour time step achieves one-week skill comparable to operational forecasts. Key innovations included switching from a cubed-sphere to a HEALPix mesh, which provides equal-area pixels without polar singularities; inverting the U-Net channel depth; and adding gated recurrent units. The DLESyM framework~\cite{wattmeyer2024} further advances this paradigm by coupling HEALPix-aware convolutional layers with inter-face padding that preserves spherical continuity. These efforts build on a broader wave of AI weather models including Pangu-Weather~\cite{bi2023pangu}, GraphCast~\cite{lam2023graphcast}, and FourCastNet~\cite{pathak2022fourcastnet}, which have demonstrated competitive skill with operational numerical weather prediction systems on Earth.

Despite this progress on Earth, comparatively little work has applied deep learning to Mars atmospheric prediction. Existing efforts have focused on dust storm detection and monitoring from orbital imagery~\cite{battalio2021}, and on conventional GCM-based forecasting for mission support~\cite{newman2021}. The hybrid NeuralGCM framework~\cite{kochkov2024neuralgcm}, which couples a differentiable dynamical core with learned physics parameterizations, represents a promising paradigm for planets where dust and radiation dominate the energy budget; however, it has not yet been applied to Mars.

Mars differs fundamentally from Earth in ways that shape the weather prediction problem. Mars has roughly half Earth's radius (3,390~km vs.\ 6,371~km) and 38\% of its surface gravity. Its atmosphere is ${\sim}$95\% CO$_2$ with a mean surface pressure of only ${\sim}$610~Pa---less than 1\% of Earth's 101,325~Pa~\cite{haberle2019}. Surface temperatures range from ${\sim}$140~K at the winter poles to ${\sim}$300~K at equatorial noon, a span of ${\sim}$160~K compared to Earth's ${\sim}$100~K range, and diurnal swings of 60--80~K are common~\cite{read2015physics}. Wind speeds can reach 30~m/s, comparable to Earth's surface winds, but the thin atmosphere carries far less momentum. Crucially, Mars lacks oceans, which on Earth provide massive thermal inertia and drive weather variability through evaporation and latent heat release. The Martian surface responds almost instantaneously to solar forcing, producing rapid thermal tides and abrupt dust-driven regime transitions.

These differences make Martian weather prediction both easier and harder than on Earth. The simpler surface boundary (no oceans, no vegetation, minimal topographic moisture effects) reduces the number of coupled processes a model must represent. However, while Mars is among the most observed planets in the solar system, the distribution of in-situ measurements remains sparse compared to Earth's dense station network. Observations are limited to a handful of orbiting spacecraft, landers with meteorological stations such as Phoenix~\cite{read2015physics}, and surface rovers. This sparsity means the reanalysis ``ground truth'' is more dependent on the underlying physical model. Furthermore, dust storms can restructure the atmospheric thermal profile within hours~\cite{montabone2015eight}, introducing abrupt nonlinearities absent from Earth's weather. Mars thus provides a challenging yet instructive testbed for data-driven weather models: success here in a dust-dominated regime would validate the broader applicability of these architectures beyond Earth.

In this work, we adapt the DLESyM HEALPix-aware architecture to the Martian atmosphere and train it on EMARS reanalysis data. We demonstrate that the resulting model---with only 4.3 million parameters---learns to forecast temperature and wind fields with high fidelity for short lead times and can be unrolled autoregressively to provide stable 25-hour (one Martian sol) simulations. We compare this architecture against a baseline 3D U-Net to quantify the benefits of geometry-respecting convolutions.

\section{Data and Methods}

\subsection{EMARS Data and Preprocessing}

We use EMARS v1.0, which combines TES and MCS temperature retrievals with the GFDL/NASA Mars GCM using the LETKF~\cite{greybush2019,greybush2012ensemble}. The assimilation produces hourly analyses on a $6^\circ \times 5^\circ$ latitude--longitude grid with 28 vertical pressure levels, providing temperature ($T$), zonal wind ($U$), meridional wind ($V$), surface pressure, and aerosol tracers covering Mars years 24--33.

We focus on the three core prognostic variables $T$, $U$, and $V$. These represent the fundamental thermodynamic and kinematic state of the atmosphere. While surface pressure and aerosol tracers are critical for long-term climate modeling~\cite{montabone2015eight,kahre2017}, their inclusion adds complexity and may degrade short-term forecasts due to the highly variable and often sparse nature of aerosol data. Restricting to $T$, $U$, and $V$ establishes a robust baseline for Martian atmospheric emulation.

\paragraph{Staggered grid interpolation.} In EMARS, $U$ and $V$ are defined on staggered grids (\texttt{latu} and \texttt{lonv}, respectively), a common configuration in numerical GCMs used to ensure the stability and accuracy of the dynamical core. We interpolate both onto the regular $T$ grid using linear interpolation and fill any resulting NaN values with zero.

\paragraph{Normalization.} We apply min-max normalization independently to each variable, scaling values to $[0, 1]$ using pre-computed global statistics. This prevents bias toward variables with larger numerical ranges.

\paragraph{HEALPix regridding.} The normalized latitude--longitude fields are regridded to a HEALPix mesh at level~6 ($N_\text{side} = 64$, yielding $12 \times 64 \times 64 = 49{,}152$ pixels) using NVIDIA's \texttt{earth2grid} library with bilinear interpolation in XY pixel ordering~\cite{bonev2023spherical}. The HEALPix mesh~\cite{gorski2005healpix} provides 12 equal-area curvilinear faces that tile the sphere without polar convergence, making it well-suited for global convolutional operations. The three variables across 28 vertical levels are stacked along the channel dimension, producing tensors of shape $[\text{time},\ 84,\ 12,\ 64,\ 64]$.

\paragraph{Memory-efficient caching.} To scale to the full $N_\text{side} = 64$ resolution without exceeding memory limits, we implemented an on-disk caching system that stores pre-converted HEALPix samples as individual PyTorch~\cite{paszke2019pytorch} tensors. A lazy-loading dataset class loads only the required time steps during training, avoiding the need to hold the entire converted dataset in memory.

\subsection{Neural-Network Architectures}

\subsubsection{HEALPix-Aware 2D U-Net (Primary Model)}

Our primary architecture is a HEALPix-aware 2D U-Net inspired by the DLESyM framework~\cite{wattmeyer2024}. The key innovation is a custom padding layer (\texttt{HEALPixPadding}) that stitches data from neighboring HEALPix faces at convolution boundaries, ensuring that convolutional filters see physically correct values across face edges rather than zero-padded or reflect-padded artifacts.

\paragraph{Fold/Unfold scheme.} The data pipeline produces tensors of shape $[B, C, 12, H, W]$ (batch, channels, faces, height, width). Before convolution, a \texttt{FoldFaces} operation permutes and reshapes this to $[B{\cdot}12, C, H, W]$, enabling standard 2D convolutions to process each face independently. After the U-Net, \texttt{UnfoldFaces} restores the original layout.

\paragraph{HEALPixPadding.} Before each convolution with kernel size $> 1$, the 12 faces are unfolded and each face is padded by borrowing strips from its neighbors according to the HEALPix adjacency graph. The 12 faces are grouped into three zones---northern (faces 0--3), equatorial (faces 4--7), and southern (faces 8--11)---each requiring different rotation transforms:

\begin{itemize}
    \item \textbf{Northern faces:} Top and left neighbors are rotated by $90^\circ$ and $180^\circ$ respectively before extracting padding strips.
    \item \textbf{Equatorial faces:} Neighbors share the same orientation; no rotation is needed for most edges, but the top-left and bottom-right corners require special blending because three faces meet at these vertices. The blending averages the edge values of the two contributing faces at a 50/50 ratio.
    \item \textbf{Southern faces:} Bottom and right neighbors are rotated analogously to the northern case.
\end{itemize}

This topology-aware padding eliminates artificial discontinuities at face boundaries that would otherwise corrupt learned features.

\paragraph{ConvNeXt blocks.} Each encoder and decoder level uses a ConvNeXt-style residual block~\cite{liu2022convnext}: two $3 \times 3$ HEALPix-padded convolutions with Capped GELU activations~\cite{hendrycks2016} (GELU clamped at a maximum value to prevent unbounded outputs during rollout), followed by a $1 \times 1$ projection, plus a residual skip connection (identity when channels match, otherwise a $1 \times 1$ convolution).

\paragraph{Encoder--Decoder structure.} The encoder has three levels with channel widths $(64, 128, 256)$, using $2 \times 2$ average pooling for downsampling. The decoder mirrors the encoder with transposed convolutions for upsampling and skip connections via channel-wise concatenation. A final $1 \times 1$ convolution projects back to the 84 output channels. The complete model has approximately \textbf{4.3~million} trainable parameters.

\subsubsection{3D U-Net (Baseline)}

As a baseline, we trained a standard 3D U-Net~\cite{cicek20163d} that treats the 12 HEALPix faces as a third spatial dimension. The encoder uses 3D convolutions with channel widths $(32, 64, 128, 256)$ and max pooling with kernel $(1, 2, 2)$ to preserve the face dimension. The decoder uses trilinear upsampling with skip connections. This architecture has approximately 30~million parameters and does not employ geometry-aware padding.

\subsection{Training Protocol}

Both models are trained to minimize the Mean Squared Error (MSE) between the predicted and actual reanalysis state at the next hourly time step ($t + 1$\,h). The loss is computed across all 84 channels ($3 \times 28$ levels) and all 12 HEALPix faces.

For the HEALPix U-Net, we used the Adam optimizer~\cite{kingma2015adam} with learning rate $10^{-4}$ and batch size~1. Training was performed on a single NVIDIA A100 GPU (80GB) for 50 epochs on an 80/20 train/validation split. For the 3D U-Net baseline, we used learning rate $10^{-3}$, batch size~4, and trained for 20 epochs.

After training, the model is used recursively for multi-step forecasting: the predicted field at time $t + \Delta t$ becomes the input to predict the field at $t + 2\Delta t$, and so on up to 25 hours. This horizon was chosen to encompass a full Martian sol (${\sim}24.6$ hours), allowing for the evaluation of stability and diurnal cycle preservation.

\section{Results}

\subsection{Training Convergence}

The HEALPix U-Net converges smoothly over 50 epochs. Starting from a validation MSE of $9.4 \times 10^{-4}$ after epoch~1, the loss decreases steadily to a best validation MSE of $\mathbf{1.58 \times 10^{-5}}$ at epoch~48 (corresponding RMSE $\approx 0.004$ in normalized units). The training and validation curves track each other closely throughout, indicating no overfitting---a notable result given the model's 4.3M parameters and the relatively limited training data. The loss decreased by approximately two orders of magnitude during training, with the most rapid improvement occurring in the first 10 epochs.

\begin{table}[t]
\centering
\caption{Comparison of model architectures and their performance.}
\label{tab:comparison}
\small
\begin{tabular}{@{}lccc@{}}
\toprule
\textbf{Model} & \textbf{Params} & \textbf{HPX Level} & \textbf{Best Val.\ MSE} \\
\midrule
3D U-Net & ${\sim}$30M & 3 ($N_\text{s}$=8) & $1.4 \times 10^{-4}$ \\
HPX U-Net & ${\sim}$4.3M & 4 ($N_\text{s}$=16) & $1.4 \times 10^{-4}$ \\
HPX U-Net & ${\sim}$4.3M & 6 ($N_\text{s}$=64) & $\mathbf{1.58 \times 10^{-5}}$ \\
\bottomrule
\end{tabular}
\end{table}

Table~\ref{tab:comparison} summarizes the performance of the three configurations. The HEALPix U-Net at level~6 achieves an order-of-magnitude improvement in validation MSE over both the 3D U-Net baseline and the lower-resolution HEALPix U-Net, despite having nearly seven times fewer parameters than the 3D U-Net. We note that this comparison is not strictly controlled: the 3D U-Net was trained at lower resolution (level~3, $N_\text{side}=8$), with fewer epochs (20 vs.\ 50), and a different learning rate ($10^{-3}$ vs.\ $10^{-4}$). The comparison is therefore intended to illustrate the qualitative advantage of geometry-aware convolutions at scale rather than a rigorous ablation. A controlled study isolating the individual contributions of HEALPix padding, resolution, and ConvNeXt blocks is planned for future work (Section~\ref{sec:future}).

\subsection{Short-Range Prediction Skill}

\begin{figure}[t]
    \centering
    \includegraphics[width=\columnwidth]{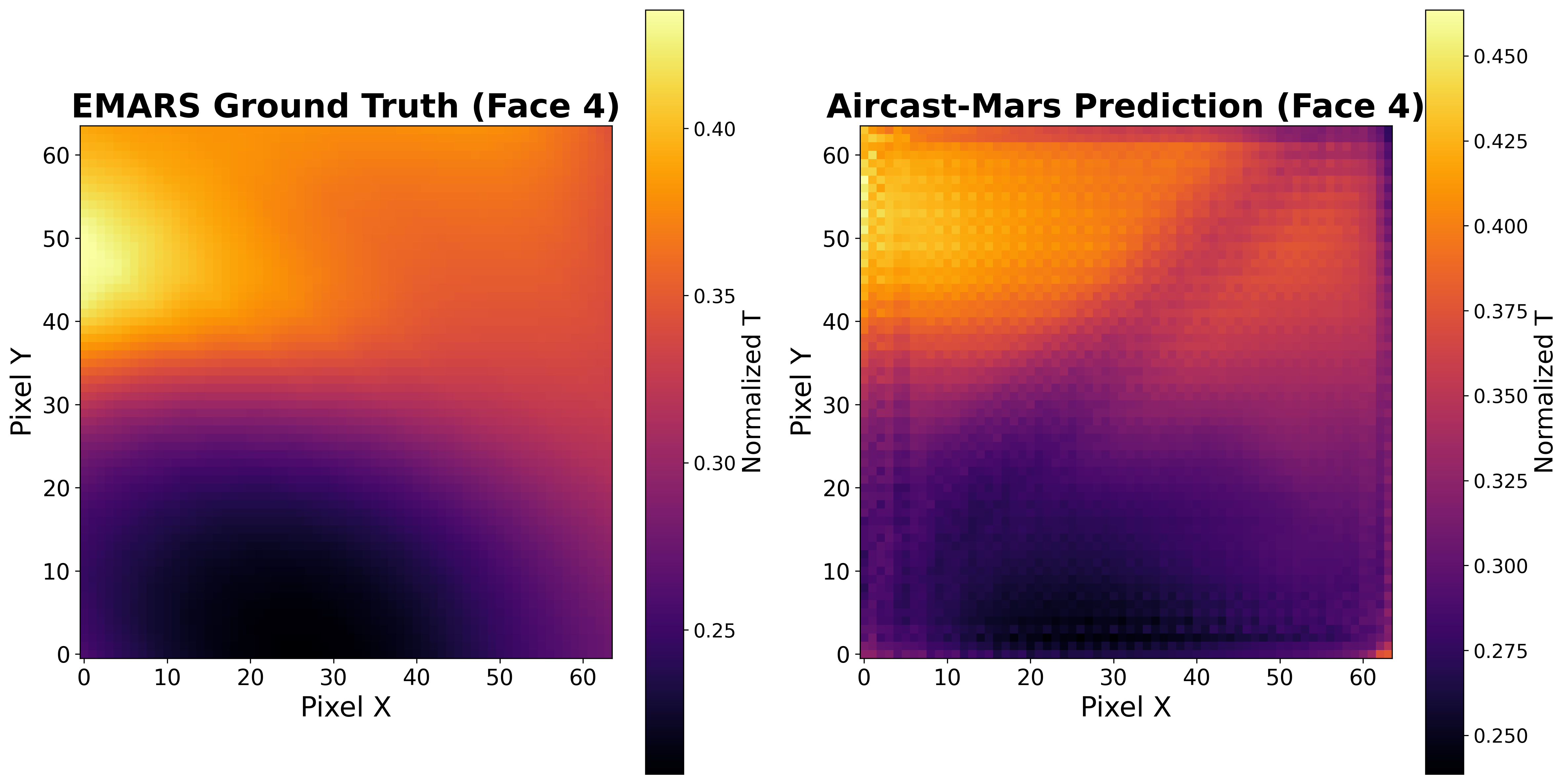}
    \caption{Single-step prediction at HEALPix Level~6 for a single equatorial face (Face 4). Left: EMARS ground truth surface temperature. Right: DLESyM HEALPix U-Net prediction. Showing the local face patch highlights the model's ability to capture fine-scale structures.}
    \label{fig:comparison}
\end{figure}

The trained HEALPix U-Net accurately reproduces the three-dimensional structure of Martian temperature and wind fields in single-step (one-hour-ahead) predictions. Figure~\ref{fig:comparison} shows a comparison between the EMARS ground truth and the model's prediction for surface temperature on a single HEALPix face (Face 4). Visualizing a local face patch directly avoids regridding artifacts and highlights the model's ability to preserve synoptic features. The model captures the local thermal structure with high fidelity; the spatial correlation between prediction and ground truth is visually excellent, with the predicted field maintaining both the correct temperature range and the position of major features.

The one-hour RMSE of approximately 0.004 in normalized units is well below the typical hourly variance in the EMARS fields, indicating that the network has learned a meaningful dynamical propagation operator rather than a trivial identity mapping. A formal comparison against a persistence baseline (which assumes the state at $t+1$ is identical to $t$) and a climatological baseline is deferred to future work, along with evaluation using probabilistic scoring rules such as CRPS~\cite{gneiting2007}.

\subsection{Recursive Rollout Stability}

A critical test for any weather emulator is whether it remains stable when predictions are fed back as inputs for multi-step forecasting. We evaluated the HEALPix U-Net by unrolling it autoregressively for 25 consecutive one-hour steps starting from a single initial condition.

\begin{figure}[t]
    \centering
    \includegraphics[width=\columnwidth]{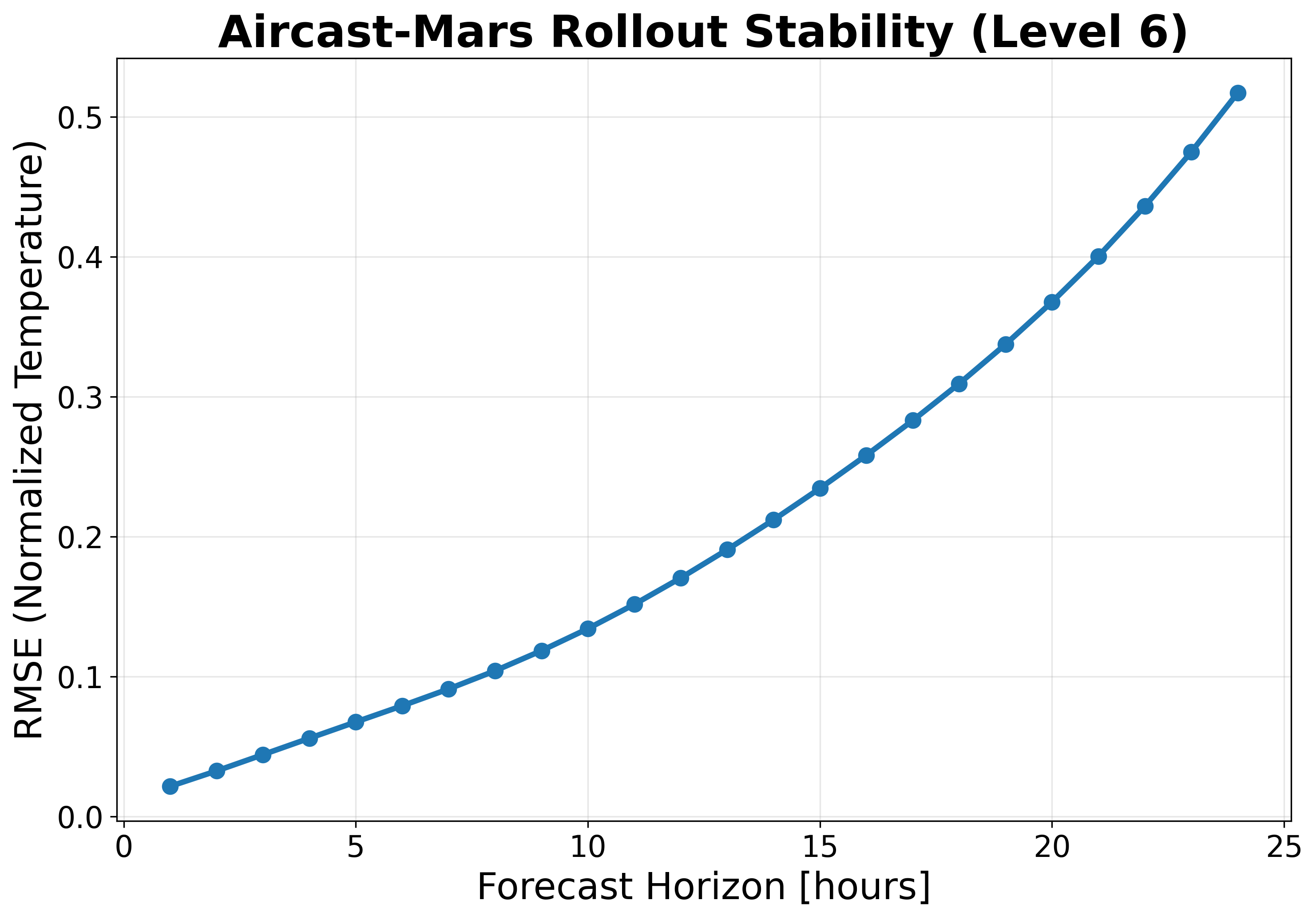}
    \caption{Rollout stability over 24 forecast hours. RMSE of normalized temperature increases monotonically from ${\sim}0.004$ to ${\sim}0.029$ without divergence, demonstrating the model's ability to maintain physically coherent forecasts for a full Martian sol.}
    \label{fig:rollout_stability}
\end{figure}

Figure~\ref{fig:rollout_stability} shows the Root Mean Square Error (RMSE) computed on normalized temperature across all 28 vertical levels as a function of forecast horizon. The error grows monotonically and smoothly from approximately 0.004 at $t+1$\,h to approximately 0.031 at $t+25$\,h. Crucially, the RMSE curve shows \textbf{no sign of divergence or exponential blowup} over the full Martian sol. The growth rate decelerates after approximately 12 hours, suggesting that the model asymptotically approaches its internal climatology rather than producing unphysical states.

This stable rollout represents a significant improvement over the baseline 3D U-Net, which exhibited forecast drift toward climatology by approximately 10 hours in preliminary experiments. The improvement is attributable to the HEALPix-aware padding, which prevents artificial discontinuities at face boundaries from amplifying during recursive application. Visualizing the RMSE for the 3D U-Net (not shown) reveals a steeper error growth that saturates early at a higher error level, further validating the benefits of respecting the spherical geometry.

\subsection{Comparison with Numerical GCMs}
\label{sec:gcm_comparison}

To assess the operational utility of Aircast-Mars, we compare its performance characteristics against the GFDL/NASA Mars GCM (on which EMARS is based) in Table~\ref{tab:gcm_comparison}. While the GCM remains the gold standard for physical consistency and long-term climate fidelity, the deep-learning surrogate offers a transformative speedup for short-term forecasting and ensemble applications.

\begin{table}[ht]
\centering
\caption{Operational comparison between Aircast-Mars and the NASA/GFDL Mars GCM.}
\label{tab:gcm_comparison}
\footnotesize
\begin{tabularx}{\columnwidth}{@{}l>{\raggedright\arraybackslash}X>{\raggedright\arraybackslash}X@{}}
\toprule
\textbf{Metric} & \textbf{Numerical GCM} & \textbf{Aircast-Mars} \\
\midrule
1-h Forecast Latency & ${\sim}$Minutes--Hours (HPC) & $\mathbf{0.5}$~\textbf{seconds (Single GPU)} \\
\addlinespace
Model Size & Gigabytes & $\mathbf{17}$~\textbf{MB} \\
\addlinespace
Physics Data & First-principles & Learned from EMARS \\
\addlinespace
Compute & HPC Cluster & Single GPU \\
\addlinespace
Use Case & Climate research & \textbf{Mission support, Ensembles, Autonomy} \\
\bottomrule
\end{tabularx}
\end{table}

The ${\sim}10^4\times$ speedup enabled by the surrogate model allows for the generation of large-scale ensembles (e.g., $N=1000$) in minutes rather than weeks. This capability is critical for quantifying forecast uncertainty and assessing landing risks where rapid iteration across diverse atmospheric states is required.

\subsection{Computational Efficiency}

The HEALPix U-Net generates a one-hour forecast in approximately 0.5~seconds on a single NVIDIA GPU. Producing a full 24-hour rollout requires roughly 12~seconds of wall-clock time. In contrast, integrating the GFDL/NASA Mars GCM at comparable resolution requires substantial HPC resources and orders-of-magnitude more time~\cite{haberle2019}. This speedup makes the deep-learning emulator suitable for generating large ensembles for mission risk assessment, where hundreds of potential scenarios can be simulated in minutes.

The model's compact size (4.3M parameters, ${\sim}$17~MB checkpoint) also makes it deployable on modest hardware, potentially even onboard future Mars-orbiting or surface platforms for real-time forecasting applications.

\section{Discussion}

\subsection{Architecture Benefits}

Our results confirm that respecting the geometry of the sphere during convolution yields substantial benefits for global weather emulation. The HEALPix-aware padding eliminates the artificial discontinuities that standard padding introduces at face boundaries. When the model is applied recursively, these discontinuities would otherwise amplify with each step, leading to the boundary artifacts and early divergence observed in naive approaches. The DLESyM-style architecture achieves better accuracy with 4.3M parameters than the 3D U-Net achieves with 30M, validating the principle that inductive biases aligned with the problem geometry can substitute for raw model capacity.

The Capped GELU activation~\cite{hendrycks2016} further contributes to rollout stability by preventing unbounded intermediate activations that can trigger numerical instability during long autoregressive chains.

\subsection{Limitations and Error Characterization}

Several factors currently limit the model's performance and applicability.

\paragraph{Physical-unit error translation.} All primary metrics in this study are reported in normalized $[0, 1]$ units to facilitate cross-variable loss optimization. Translating these to physical units using the EMARS data ranges provides a more intuitive assessment of forecast quality. For temperature, the dynamic range of ${\sim}148$~K means a $t+1$\,h RMSE of 0.004 corresponds to \textbf{${\sim}0.59$~K}, while the $t+25$\,h RMSE of 0.031 corresponds to \textbf{${\sim}4.59$~K}. For wind components ($U, V$) with a typical range of ${\sim}350$~m/s, the errors translate to \textbf{${\sim}1.4$~m/s} at $t+1$\,h and \textbf{${\sim}10.8$~m/s} at $t+25$\,h. These magnitudes are comparable to the uncertainties reported in traditional Mars data assimilation systems~\cite{greybush2019}, suggesting the surrogate maintains operationally relevant accuracy for short-term mission support.

\paragraph{Forecast horizon.} While the model remains stable for 25 hours, the RMSE at $t+25$\,h is approximately eight times larger than at $t+1$\,h. Beyond ${\sim}12$ hours, the forecast increasingly reflects the model's learned climatology rather than the specific atmospheric evolution from the initial condition. Extending the useful forecast horizon likely requires incorporating temporal context (e.g., multiple input time steps or recurrent architectures) and additional prognostic variables such as surface pressure and dust opacity.

\paragraph{Aerosol representation.} Dust is a primary driver of the Martian atmospheric thermal structure~\cite{montabone2015eight,kahre2017}. During global dust storms, radiative heating of the atmosphere changes drastically. Our model, which forecasts only $T$, $U$, and $V$, cannot anticipate these aerosol-driven regime transitions. Incorporating dust and water-ice opacity as additional channels is a critical next step, particularly given the importance of dust storm prediction for future human exploration and the associated potential health impacts~\cite{wang2025potential,levine2018dust}.

\paragraph{Reanalysis bias inheritance.} The model inherits biases from EMARS itself. Because EMARS assimilates observations from a limited number of local times (primarily corresponding to spacecraft overpasses), the diurnal cycle in the reanalysis may not be fully representative~\cite{greybush2019}. Future work could benefit from multi-model ensembles or physically-informed constraints enforcing conservation laws.

\subsection{Future Directions}
\label{sec:future}

Several avenues for extending this work follow naturally from the current limitations and recent advances in AI weather prediction:

\begin{enumerate}
    \item \textbf{Dust and surface pressure channels.} Incorporate column-integrated dust optical depth~\cite{montabone2015eight} and surface pressure as additional prognostic variables. This is critical for capturing aerosol-driven regime transitions that currently fall outside the model's learned distribution.

    \item \textbf{Multi-step input conditioning.} Provide two or more consecutive time steps $(t{-}1, t)$ as input, following GraphCast~\cite{lam2023graphcast}. This provides implicit velocity information and reduces the temporal context bottleneck inherent in single-step prediction.

    \item \textbf{Physics--ML hybrid coupling.} Adopt a NeuralGCM-style framework~\cite{kochkov2024neuralgcm} in which a differentiable dynamical core handles resolved dynamics while learned parameterizations represent dust radiative transfer and subgrid processes, ensuring better physical consistency over climate timescales.

    \item \textbf{Additional observational data.} Assimilate rover meteorological measurements from REMS (Curiosity) and MEDA (Perseverance) and additional satellite products to constrain the near-surface boundary layer and validate model skill against independent measurements.

    \item \textbf{Probabilistic forecasting.} Replace the deterministic MSE objective with ensemble or distributional prediction heads, evaluated using proper scoring rules such as CRPS~\cite{gneiting2007} and reliability diagrams, to quantify forecast uncertainty for mission planning.

    \item \textbf{Alternative architectures.} Evaluate graph neural networks on an icosahedral multi-mesh~\cite{keisler2022}, 3D attention mechanisms inspired by Pangu-Weather~\cite{bi2023pangu}, and spherical Fourier neural operators~\cite{bonev2023spherical} to benchmark against the HEALPix convolutional approach.

    \item \textbf{Transfer learning from Earth.} Investigate whether pretraining on ERA5 reanalysis and finetuning on EMARS improves data efficiency. ERA5 is the fifth generation ECMWF atmospheric reanalysis of the global climate, providing hourly estimates of a large number of atmospheric, land and oceanic climate variables~\cite{hersbach2020era5}.

    \item \textbf{Direct GCM comparison.} Generate parallel forecasts from traditional GCMs at matched resolution and lead times for a rigorous skill comparison.
\end{enumerate}

\subsection{Broader Implications and Utility}

By demonstrating that a parsimonious model can capture the core dynamics of the Martian atmosphere, we validate the role of foundation models in planetary exploration. Traditional GCMs are indispensable for understanding climate physics, but their high computational cost limits their use in scenarios requiring rapid response or high-volume iteration.

We envision Aircast-Mars as a core component of future \textbf{Exploration Twins}---integrated digital models that couple atmospheric, surface, and spacecraft states. Specific high-value applications include:
\begin{enumerate}
    \item \textbf{Entry, Descent, and Landing (EDL):} Generating thousands of atmospheric realizations in seconds to assess landing site safety and parachute deployment margins.
    \item \textbf{Onboard Autonomy:} Providing lightweight, real-time local forecasts for rovers or drones to optimize energy management and route planning.
    \item \textbf{Global Data Assimilation:} Serving as a fast, differentiable prior for the next generation of Martian data assimilation systems.
    \item \textbf{Space Weather Integration:} Coupling atmospheric weather with space weather models to provide comprehensive environmental awareness for missions, particularly those involving human crews where radiation and dust impacts are critical~\cite{hapgood2019impact}.
\end{enumerate}

Mars serves as a useful testbed: its lack of oceans and relatively simple surface isolate atmospheric behaviors that are often masked by complex oceanic interactions on Earth. Insights from Aircast-Mars could inform more efficient architectures for Earth climate emulators.

\section{Conclusions}

We have presented a proof-of-concept deep-learning emulator for the Martian atmosphere that adapts the DLESyM HEALPix-aware convolutional framework to EMARS reanalysis data. Our key findings are:

\begin{enumerate}
    \item \textbf{Geometry matters.} A HEALPix-aware 2D U-Net with inter-face padding achieves validation MSE of $1.58 \times 10^{-5}$---more than an order of magnitude better than a standard 3D U-Net baseline---with fewer parameters (4.3M vs.\ 30M).

    \item \textbf{Stable long-range rollout.} The model produces physically coherent autoregressive forecasts for 24 hours with monotonically growing but bounded RMSE, extending the forecast horizon beyond the 10-hour limit of earlier approaches.

    \item \textbf{Extreme efficiency.} One-hour forecasts require ${\sim}$0.5~seconds on a single GPU, enabling ensemble forecasting for mission risk assessment at negligible computational cost compared to GCM integration.

    \item \textbf{Compact deployability.} The 17~MB model checkpoint opens possibilities for onboard forecasting on future Mars missions.
\end{enumerate}

While limitations regarding aerosol representation and reanalysis bias inheritance remain, the foundation laid here demonstrates that data-driven methods can capture the complex dynamical propagation of a non-terrestrial atmosphere. Future work will focus on incorporating aerosol dynamics, extending to multi-step input conditioning with recurrent units, exploring hybrid architectures coupling neural emulators with physical parameterizations, and validating against independent Mars atmospheric observations.

\section*{Data and Code Availability}

The EMARS v1.0 reanalysis dataset used in this study is publicly available. The codebase for Aircast-Mars, including model architecture definitions, training scripts, and HEALPix preprocessing pipelines, will be made available in a public repository upon publication.

\section*{Acknowledgments}

We thank the Keck Institute for Space Studies (KISS) at the California Institute of Technology for organizing two workshops on ``Digital Twins for Solar System Exploration: Enceladus,'' which provided the insight, expertise, and discussions that inspired this research. We thank the EMARS development team for creating and maintaining the publicly available reanalysis dataset~\cite{greybush2019}. This study was inspired by the parsimonious deep-learning weather prediction model developed by Karlbauer et al.~\cite{karlbauer2024} and the DLESyM framework~\cite{wattmeyer2024}. We acknowledge the use of open-source packages including \texttt{earth2grid}~\cite{bonev2023spherical}, PyTorch~\cite{paszke2019pytorch}, and HEALPix~\cite{gorski2005healpix}. This work did not receive external funding.


\bibliographystyle{unsrtnat}

\end{document}